# Thin films of PS/PS-b-PNIPAM and PS/PNIPAM polymer blends with tunable wettability

Maria Kanidi[1,2], Aristeidis Papagiannopoulos[1], Athanasios Skandalis[1], Maria Kandyla[1,*], and Stergios Pispas[1]

[1]*Theoretical and Physical Chemistry Institute, National Hellenic Research Foundation, 48 Vasileos Constantinou Avenue, 11635 Athens, Greece*

[2]*Department of Material Science, University of Patras, University Campus, 26504 Rio, Greece*

Correspondence to: Maria Kandyla (E-mail: *kandyla@eie.gr*)

((Additional Supporting Information in the online version of this article.))

**ABSTRACT**

We develop thin films of blends of polystyrene (PS) with the thermoresponsive polymer poly(N-isopropylacrylamide) (PNIPAM) (PS/PNIPAM) and its diblock copolymer polystyrene-b-poly(N-isopropylacrylamide) (PS/PS-b-PNIPAM) in different blend ratios and we study their surface morphology and thermoresponsive wetting behaviour. The blends of PS/PNIPAM and PS/PS-b-PNIPAM are spin-casted on flat silicon surfaces with various drying conditions. The surface morphology of the films depends on the blend ratio and the drying conditions. The PS/PS-b-PNIPAM films do not show an increase of their water contact angles with temperature, as it is expected by the presence of the PNIPAM block. All PS/PNIPAM films show an increase in the water contact angle above the lower critical solution temperature of PNIPAM, which depends on the ratio of PNIPAM in the blend and is insensitive to the drying conditions of the films. The difference between the wetting behaviour of PS/PS-b-PNIPAM and PS/PNIPAM films is due to the arrangement of the PNIPAM chains in the film.

**KEYWORDS:** PNIPAM, PS-b-PNIPAM, thermoresponsive wettability, blends, thin films, diblock copolymer, spin coating

## INTRODUCTION

Wettability is an essential property of solid surfaces, associated with the surface chemistry and topography.[1] The control of wettability is related to many important technological applications such as self-cleaning surfaces for satellite dishes and solar energy panels, waterproof textiles, microfluidics, coatings for boats, metal refining, *etc*.[1,2] Additionally, wettability control is essential for biological applications, such as controlled drug delivery, cell encapsulation, enzyme immobilization, cell adhesion, and biosensor development, among others.[4] In general, the wetting behaviour of a surface is divided into four states, depending on the water contact angle with the surface ($\vartheta$). Hydrophilicity ($\vartheta$ = 10°–90°) and hydrophobicity ($\vartheta$ = 90°–150°) are the two main states of the wetting behaviour, while superhydrophilicity corresponds to $\vartheta$ = 0°–10° and superhydrophobicity to $\vartheta$ = 150°–180°. Recently, smart surfaces have been developed with a switchable wetting behaviour between hydrophilicity and hydrophobicity by external stimuli, such as temperature, pH, light, *etc*.[2] Polymer films are often used as stimuli-responsive materials, which allow simultaneous control of the surface chemistry and topography, thus forming smart surfaces with controllable wettability.[5]

A well-known stimuli-responsive polymer is poly(N-isopropylacrylamide) (PNIPAM), the solubility of which changes in response to temperature. At the lower critical solution temperature (LCST) of 32 °C, PNIPAM in aqueous solution undergoes a phase transition from a soluble (below 32 °C) to an insoluble state.[6] Below the LCST, the PNIPAM chains are hydrophilic due to intermolecular H-bonding with water molecules.[5] On the other hand, above the LCST, the PNIPAM chains form a compact and collapsed conformation, which prohibits the interaction of the hydrophilic groups C=O and N–H with water.[5,7] Because the LCST of PNIPAM is close to the human body temperature, PNIPAM in hydrogel, solution, and nanoparticles is extensively investigated for biomedical applications, such as controllable drug release, tissue regeneration, cell-culture substrates, and filtration membranes, among others.[8-12]

Additionally, the combination of PNIPAM with other polymers in block copolymer systems has been used for the development of smart surfaces with a controllable wetting behaviour.[12-17] The advantage of block copolymer surfaces is the combination of the different properties of the blocks and the formation of new nanopatterned film morphologies that depend on many factors, such as the temperature and solvent conditions, electric and magnetic fields, the substrate topography, and chemical composition.[18] A well-studied and stable monomer to copolymerize with PNIPAM is polystyrene (PS). Approaches to create surfaces with tunable wettability have developed polystyrene-b-poly(N-isopropylacrylamide) (PS-b-PNIPAM) on silicon and glass substrates via the 'grafting from' mechanism.[12,16,17] The growth of controllable block copolymer brushes is achieved by the grafting reaction that proceeds by polymerization from the surface. Block copolymers that are developed by using the 'grafting from' method are often synthesized with atom transfer radical polymerization (ATRP) reactions and anionic and cationic grafting techniques, which allow the formation of stable layers of copolymers and the variation of the polymer composition, the topology of the formed layer, and its functionalities.[19] However, a major drawback of grafting is that it requires elaborate equipment and metal catalysts, which need to be removed by complicated and costly methods at the end of the polymerization process.

Polymers and block copolymers can also be blended, increasing the possible morphologies of the resulting thin films and the combinations of different physical and chemical properties. Block copolymers, blends of homopolymers, and blends of a block copolymer and a homopolymer create complex film morphologies that usually present micro- or macrophase separation on the surface. In block copolymers, microphase separation can occur between the copolymer blocks, forming nanoscale structures and minimizing the interfacial area and chain stretching.[18-22] In homopolymer blends, macrophase separation can occur between the homopolymers.[21,22] In blends of a block copolymer and a homopolymer, the molecular weight of the homopolymer and the diblock copolymer affect the phase separation.[22,23] For a homopolymer molecular weight much higher than the molecular weight of the block copolymer, both micro- and macrophase separation can occur between the copolymer blocks and the copolymer and homopolymer, respectively.[22] On the other hand, microphase separation dominates when the homopolymer molecular weight is much lower than the block copolymer.[22] Many parameters affect the phase separation in films of polymer blends, such as the blend components and their interaction with the substrate, the film thickness, the annealing conditions, and the casting solvent.[24,25] Phase separation can induce micro- and nano-domains on the film surface, creating a complex morphology, and it can determine the surface topography. Consequently, phase separation affects the wettability of thin films, as it has been observed that surface roughness can enhance the wettability of solid surfaces due to a higher surface energy.[3]

In this work, we develop thin films of blends of PS with either PNIPAM, PS/PNIPAM, or its diblock copolymer PS-b-PNIPAM, PS/PS-b-PNIPAM, in various blend ratios and we study their surface morphology and thermoresponsive wetting behaviour. The blends of PS/PNIPAM and PS/PS-b-PNIPAM are spin-casted on flat silicon surfaces and dried under various conditions. Spin coating is a widely used technique for the development of homogeneous thin films over large areas, offering reproducibility and control of the process.[26] In contrast to grafting techniques, spin coating is simple, cost-effective, and does not require the use of catalysts. Although there are several studies on the wetting behaviour of PS-b-PNIPAM grafted films, [12,16,17] there are no studies on the thermoresponsive wetting behaviour of spin-casted films of PS/PS-b-PNIPAM and PS/PNIPAM blends. The PS/PS-b-PNIPAM films do not show an increase of their contact angles with temperature, regardless of the blend ratio and drying conditions. On the other hand, all PS/PNIPAM films present an increase in the water contact angle above the LCST of PNIPAM. The homopolymer blends are promising for the development of smart surfaces by simple, cost-effective processes, suitable for applications that require wettability control.

**EXPERIMENTAL**

Materials. Polymers utilized in this study were synthesized in-house. In particular, polystyrene ($PS_{1488}$) homopolymer ($M_w$=155,000, $M_w/M_n$=1.05) was synthesized by anionic polymerization. Poly(N-isopropylacrylamide) ($PNIPAM_{265}$) homopolymer ($M_w$=30,000, $M_w/M_n$=1.16) and diblock polystyrene-b-poly(N-isopropylacrylamide) ($PS_{893}$-b-$PNIPAM_{106}$) copolymer (Mw=18,300, $M_w/M_n$=1.26, 60%wt PS) were synthesized by RAFT polymerization. Solvent THF utilized for film preparation was of analytical grade and was used without further purification. For contact angle measurements water freshly distilled from an all glass distillation apparatus was used.

Film preparation. Blends of $PS_{1488}$/$PS_{893}$-b-$PNIPAM_{106}$ and $PS_{1488}$/$PNIPAM_{265}$ were prepared in blend ratios of 50/50 and 75/25 for each formula. They were diluted in THF, consisting of 1 wt% polymer solution and stayed overnight. Silicon wafers ((100), p-type, ρ = 1–10 Ω·cm, thickness 500 ± 25 μm) were cleaned in an ultrasonic bath of acetone and methanol for 15 min each. Blend films were spin-casted onto clean silicon wafers at 3000 rpm for 30 s. After spin coating, we followed three different drying protocols for each polymer formula and blend ratio: a) the films were dried in ambient conditions at room temperature (not annealed), b) the films were annealed at 100 °C for 5 min, and c) at 100 °C for 2 hours. Reference polymer films of PS and PS-b-PNIPAM were also spin-casted on silicon substrates and used for comparison with the blend films. All films were characterized by scanning electron microscopy (SEM), optical microscopy, and micro-Raman spectroscopy. Film thickness was measured by profilometry. Films of PS/PNIPAM blends are 100 ± 30 nm thick and films of PS/PS-b-PNIPAM blends are 160 ± 50 nm thick. The range of the film thickness is attributed to the annealing procedure, because the annealed films of both kinds of blends are thinner than the not annealed ones, especially the films after 2 hours of annealing. The average size of the features on the film surface was determined by using an image processing program (ImageJ), measuring three times the size of one hundred features of each film in SEM images. Especially for the (not annealed) 75/25 PS/PNIPAM film, the average size was determined by measuring three times the size of twenty features.

Micro-Raman spectroscopy setup. Raman spectra were acquired with a Renishaw inVia Reflex Raman microscope, equipped with a Peltier-cooled charge coupled device (CCD) and a motorized xyz microscope stage with a lens of magnification ×100, in a backscattering geometry. The 514.5 nm line of an argon laser was used for excitation. Together with the rest of the system configuration (grating, slit width, CCD partition) this results in a spectral resolution of ~1 cm$^{-1}$. The laser beam was focused on the sample to a spot diameter 1–2 µm and the excitation laser power was 0.10 mW.

Contact angle measurements. A water drop of 5 µL was deposited on the surface of each film to measure the contact angle at room temperature (25 °C) and at 45 °C, using a constant temperature control base. Images of water drops on the surface were obtained by a 2.0MP 500x USB Digital Microscope and analyzed by standard MATLAB functions. In detail, the images were transformed to grayscale and subsequently the drop edge profile was determined. The profile was numerically analyzed in polar coordinates $(r, \theta)$. The origin of the coordinates was chosen on the intersection between the apparent perpendicular axis of symmetry of the drop and the solid/liquid interface. The profiles $r(\theta)$ obtained this way vary smoothly with $\vartheta$ and were easily modeled by fourth-order polynomials. The contact angles were extracted from the derivatives (transformed back to Cartesian coordinates) on the left and right-side contacts of the drop with the surface. Typically, the deviation between left and right contact angles was less than 2º. The average of the two values was reported as the measured contact angle. For each film, water contact angles were measured three times.

**RESULTS AND DISCUSSION**
**3.1 Blends of PS/PS-b-PNIPAM**

Scanning electron micrographs (SEM) of spin-casted films of blends of PS/PS-b-PNIPAM in weight ratios of 75/25 and 50/50, as well as of a pure PS-b-PNIPAM film on flat silicon substrates, before and after the water contact angle measurements, are presented in Figure 1

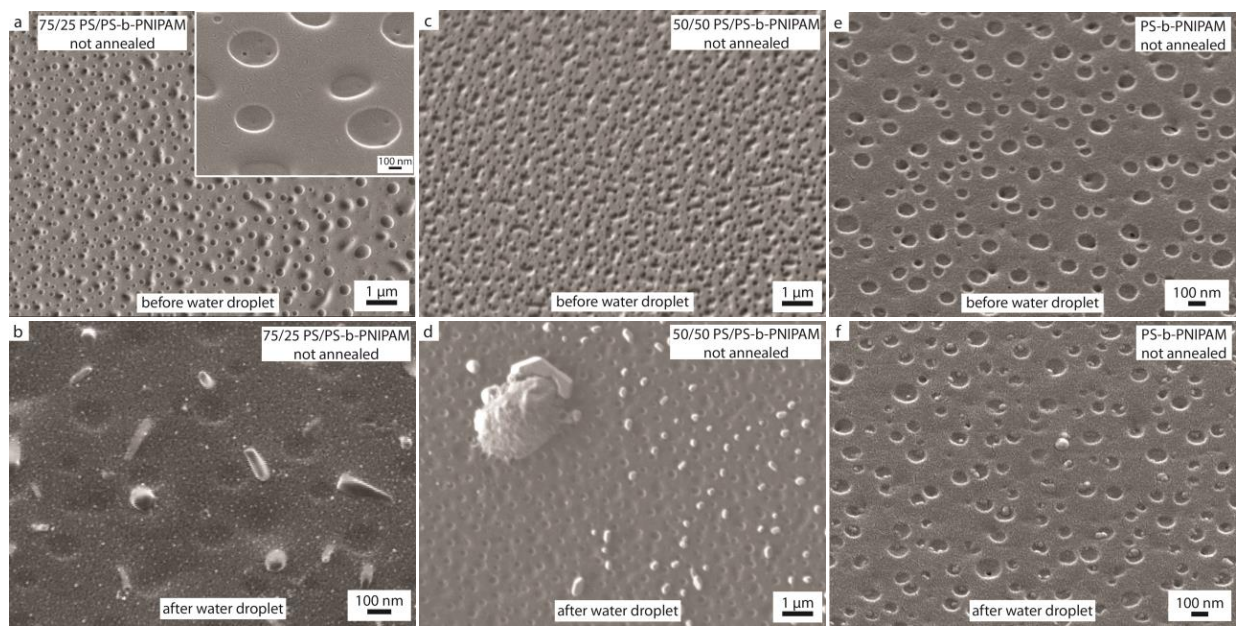

FIGURE 1. (a) Scanning electron micrographs (SEM) of a 75/25 PS/PS-b-PNIPAM film, not annealed, before the water contact angle measurement at top view with an inset of high magnification at side (45°) view and (b) after the water contact angle measurement at side (45°) view. (c) SEM of a 50/50 PS/PS-b-PNIPAM film, not annealed, before the water contact angle measurement at side (45°) view and (d) after the water contact angle measurement at top view. (e) SEM of a pure PS-b-PNIPAM film, not annealed, before the water contact angle measurement at side (45°) view and (f) after the water contact angle measurement at side (45°) view.

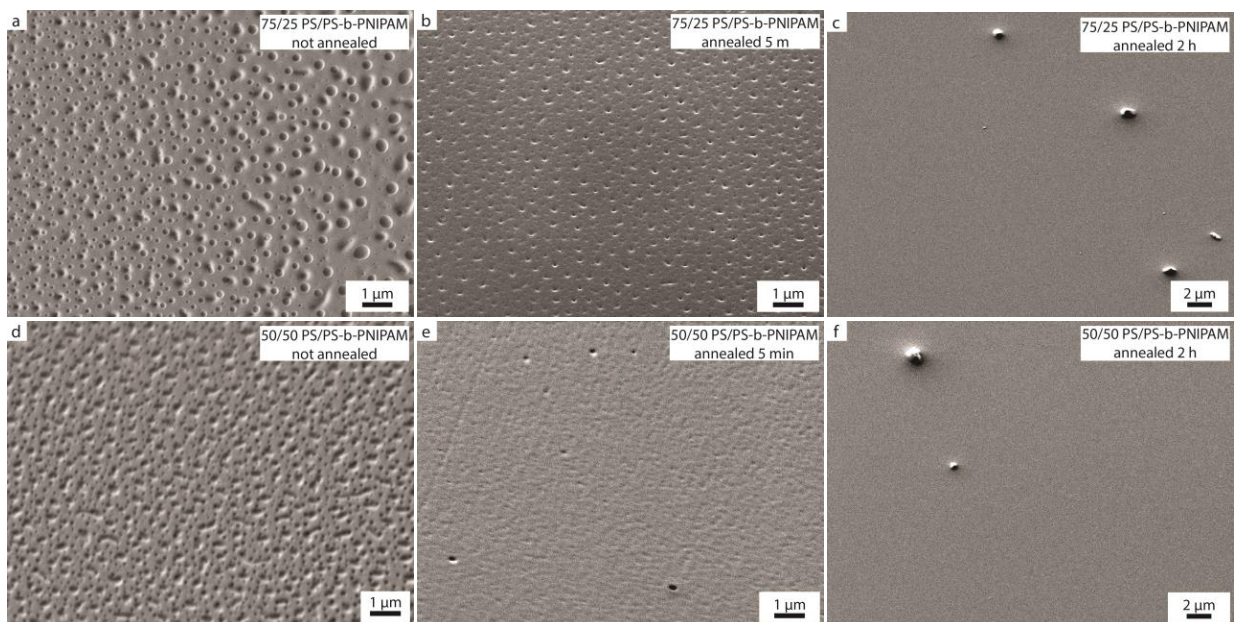

FIGURE 2. Scanning electron micrographs (SEM) of a 75/25 PS/PS-b-PNIPAM film (a) not annealed at top view, (b) annealed at 100 °C for 5 min at side (45°) view, and (c) annealed at 100 °C for 2 hours at side (45°) view. SEM at side (45°) view of a 50/50 PS/PS-b-PNIPAM film (d) not annealed, (e) annealed at 100 °C for 5 min, and (f) annealed at 100 °C for 2 hours.

Figure 1a and b show a PS/PS-b-PNIPAM film with a 75/25 ratio, not annealed, before and after the contact angle measurement, respectively, with a morphology of holes on a continuous background with an average diameter of 221 ± 83 nm. A PS/PS-b-PNIPAM film, not annealed, with a 50/50 ratio, presents small and dense holes of 111 ± 23 nm average diameter in Figs. 1c and d. The pure PS-b-PNIPAM film consists also of an array of holes with an average diameter of 76 ± 21 nm (Figs. 1e and f). All the films of PS/PS-b-PNIPAM and PS-b-PNIPAM are stable, as the water droplet for the contact angle measurement does not change their surface morphology, leaving only some impurities on the surface (Figs. 1b, d, and f). Comparing Figs. 1a, c, and e, we observe that the blend ratio affects the surface morphology of the films, resulting in different diameters and densities of holes. Increasing the ratio of the diblock copolymer PS-b-PNIPAM in the blend, the diameter and spacing between the holes decrease. The nanometer-scale structure of the pure PS-b-PNIPAM film may be associated with microphase separation, occurring between the two blocks.[27] In the case of PS/PS-b-PNIPAM films, most likely macrophase separation occurs because the molecular weight of PS is higher than the molecular weight of PS-b-PNIPAM.[22] Therefore, we assume that the observed morphology in the SEM images is associated with macrophase separation, considering the larger length scale.

Figure 2 shows the effect of annealing on PS/PS-b-PNIPAM films of 75/25 and 50/50 blend ratios. For effective annealing, the annealing temperature must be close to the glass-transition temperature ($T_g$) of both blocks (PS and PNIPAM) in order to provide the required mobility for the polymer chains to rearrange.[18] Even though $T_{g(PS)}$ is 100 °C and $T_{g(PNIPAM)}$ is 135 °C, we set the annealing temperature no higher than 100 °C, in order to avoid polymer degradation. Figures 2a, b, and c show SEM images of the 75/25 film, not annealed and annealed at 100 °C for 5 min and 2 hours, respectively. We observe that the holes on the surface of the 75/25 film persist for the most part after 5 min of annealing (Fig. 2b), even though they become smaller in size. After 2 hours of annealing (Fig. 2c), the surface of the 75/25 film becomes completely smooth without holes. For the 50/50 film, the holes on the surface are reduced for the most part after 5 min of annealing (Fig. 2e) and after 2 hours of annealing they are eliminated completely (Fig. 2f). The fact that the 75/25 PS/PS-b-PNIPAM film requires more annealing time to become completely smooth is explained by the increased ratio of PS, which has a higher $T_g$, close to the annealing temperature. As expected, in all cases thermal annealing offers the required energy for chain rearrangement and leads to more ordered structures in the block copolymer films.[28]

Water contact angle measurements at 25 °C and 45 °C reveal the wetting behaviour of PS/PS-b-PNIPAM films of 75/25 and 50/50 blend ratios and the pure PS-b-PNIPAM film. All films are hydrophilic with contact angles more than 70° (Fig. 3). The highest contact angle is close to 90° for the annealed 75/25 films. Neither the PS/PS-b-PNIPAM blend films nor the pure PS-b-PNIPAM film show an increase of the contact angle above 32 °C, as it would have been expected due to the thermoresponsive wettability of PNIPAM. This result can be explained by the fact that the block of PNIPAM enriches the layer close to the silicon substrate and not the upper layer of the polymeric film surface, which inhibits the manifestation of the PNIPAM thermoresponsivity.[29,30] This arrangement is not reversed by annealing, as we observe in Fig. 3 that the annealed films show the same wetting trend as their not annealed counterparts, even though annealing severely affects the surface morphology (Fig. 2). However, we observe that the contact angles of the PS/PS-b-PNIPAM films are not entirely the same below and above 32 °C. Most of the PS/PS-b-PNIPAM films present small changes in their water contact angles below and above 32 °C, within the measurement error. Three films present changes that exceed the measurement error, especially the 2-hour annealed 75/25 and 50/50 PS/PS-b-PNIPAM films and the not annealed 50/50 PS/PS-b-PNIPAM film. Although there are small changes in the contact angles upon heating, the PS/PS-b-PNIPAM films do not respond systematically to the stimulus of temperature in order to allow reliable conclusions.

### 3.2 Blends of PS/PNIPAM

In Figure 4 we present SEM images of spin-casted films of blends of PS/PNIPAM of 75/25 and 50/50 blend ratios on flat silicon substrates, before and after the water contact angle measurements. Figures 4a and b show the 75/25 PS/PNIPAM film, not annealed, shows a continuous surface with circular features (average diameter 8 ± 2 μm), where the film is dewetted, on a homogeneous background. Actually, there are areas where the film is dewetted and areas where the film is not completely dewetted, creating random domains as indicated by the colored arrows. The surface morphology with random domains is attributed to the synergy of the dewetting mechanism with phase separation, which depends on parameters such as molecular weight of the components, surface energy of the substrate, relative solubility of the polymers in the solvent, and the film thickness, among others.[31,32]

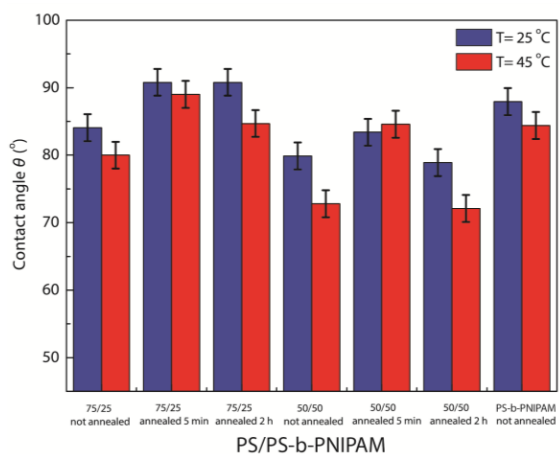

FIGURE 3. Bar graph of water contact angles, measured at 25 °C and 45 °C, on a pure PS-b-PNIPAM film and on PS/PS-b-PNIPAM films of 75/25 and 50/50 blend ratios (not annealed, annealed at 100 °C for 5 min, and annealed at 100 °C for 2 hours).

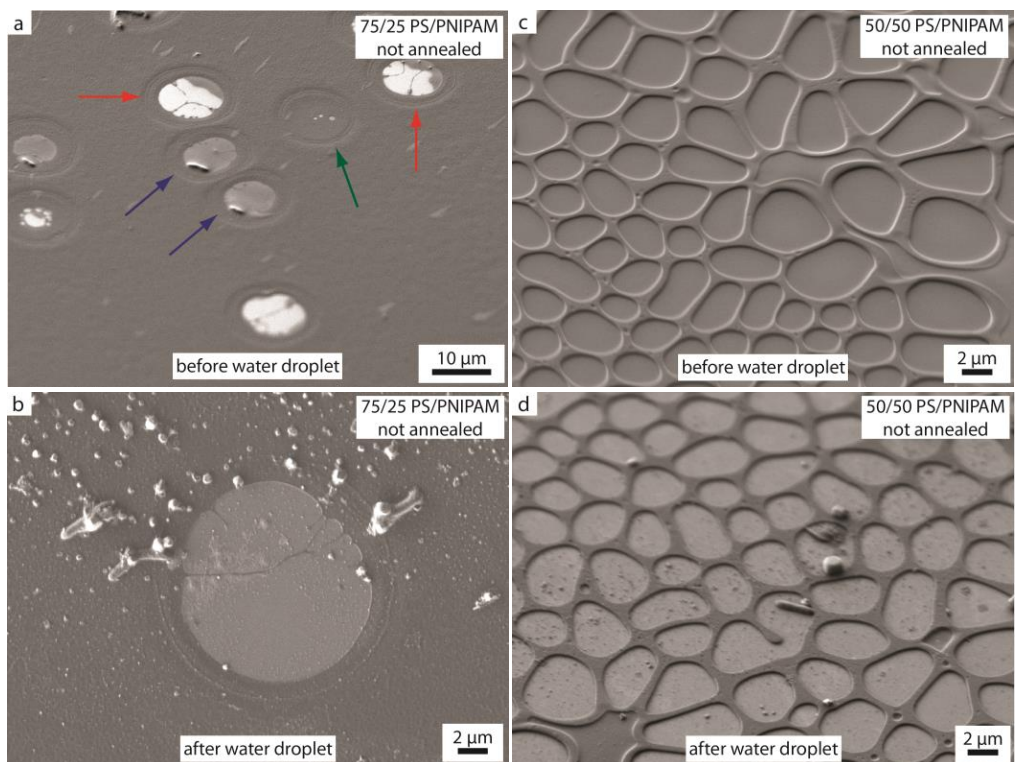

FIGURE 4. (a) Scanning electron micrographs (SEM) of a 75/25 PS/PNIPAM film, not annealed, before the water contact angle measurement at side (45°) view and (b) after the water contact angle measurement at top view. In Fig. 4a, red arrows indicate dewetted areas, blue arrows and the green arrows indicate areas that are not completely dewetted. (c) SEM of a 50/50 PS/PNIPAM film, not annealed, before the water contact angle measurement at side (45°) view and (d) after the water contact angle measurement at side (45°) view.

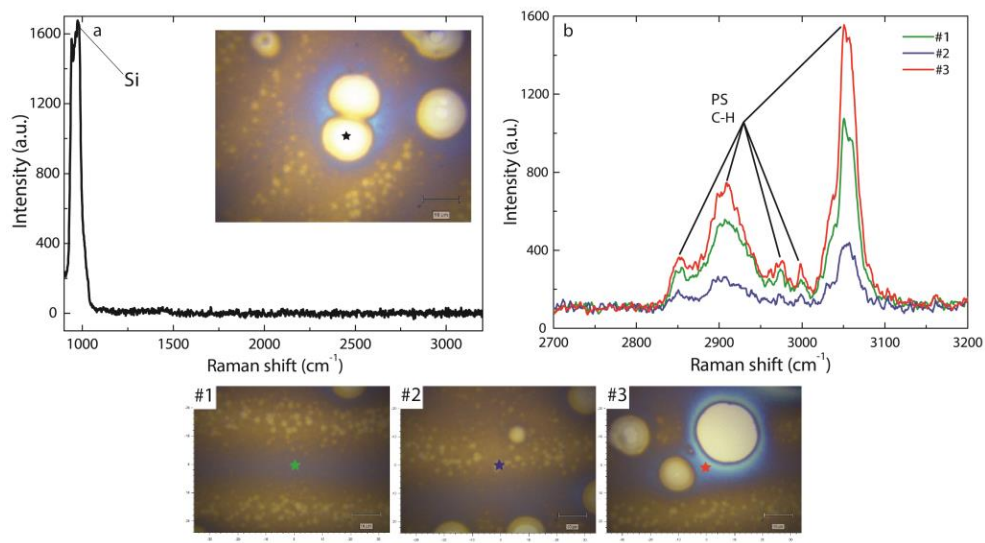

FIGURE 5. Raman spectra, measured with 514 nm excitation wavelength, of a 75/25 PS/PNIPAM film (a) inside the dewetted areas and (b) on three different locations of the film outside the dewetted areas. The spectra correspond to the locations marked with similar colored stars on the optical micrographs.

In the case of 50/50 blend ratio, Figs. 4c and d show the 50/50 PS/PNIPAM film, not annealed, which forms an interconnected network with isolated dewetted areas of an average diameter of 3 ± 0.9 µm. The formation of the interconnected network morphology is attributed to the coexistence of dewetting and phase separation. Dewetting is more pronounced in the 50/50 PS/PNIPAM film than in the 75/25 film due to the higher ratio of PNIPAM. Similar to Figs. 1b, d, and e, the water droplet of the contact angle measurement deposits some impurities on the film surface but does not alter the film morphology (Figs. 4b and d). Although the morphology of the PS/PNIPAM films depends on the blend ratio similarly to the PS/PS-b-PNIPAM films, the features on the surface of PS/PNIPAM are micrometric and not nanometric, which is the case for PS/PS-b-PNIPAM films. The different composition of the polymer blends and consequently their molecular weights are the reasons for the domination of the phase separation effect on the surface morphology of PS/PS-b-PNIPAM films in contrast to the surface morphology of PS/PNIPAM films, which is dominated by dewetting. The coexistence of phase separation and dewetting in immiscible films is very difficult to quantify and only recently certain attempts have been made to explain these phenomena.[33]

To investigate further the topography of the PS/PNIPAM films we perform micro-Raman spectroscopy. In the case of the 75/25 PS/PNIPAM film, Raman spectra inside the circular features (Fig. 5a) show only the overtone (ca. 963 cm$^{-1}$) of the silicon optical phonon peak (520.5 cm$^{-1}$) and no PS or PNIPAM peaks, confirming that the circular features are holes where the film is dewetted. Figure 5b shows Raman spectra on three different locations of the PS/PNIPAM surface outside the dewetted areas, which are indicated by the corresponding stars in the optical micrographs. All the peaks correspond to the vibrational modes of C-H for PS chains in the spectral region 2800–3200 cm$^{-1}$ (Fig. 5b). Specifically, the vibrational mode of C-H for PS chains appears at 2852 cm$^{-1}$, 2907 cm$^{-1}$, 2976 cm$^{-1}$, 2999 cm$^{-1}$, and 3054 cm$^{-1}$.[34] We verify the presence of these peaks by Raman spectroscopy on solid PS (Figs. S1 and S2 in Supplementary Information).

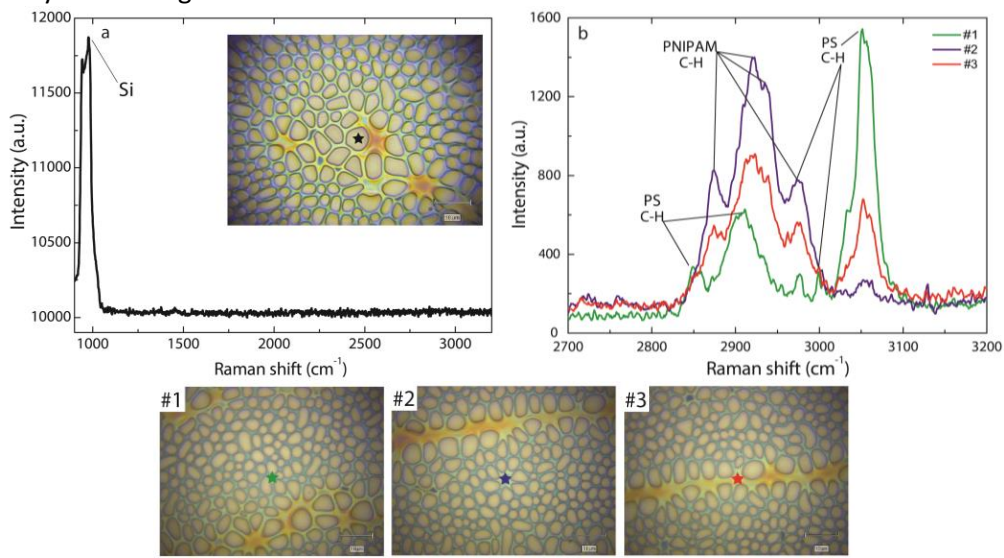

FIGURE 6. Raman spectra, measured with 514 nm excitation wavelength, of a 50/50 PS/PNIPAM film (a) inside the dewetted areas and (b) on three different locations of the PS/PNIPAM interconnected network. The spectra correspond to the locations marked with similar colored stars on the optical micrographs.

The absence of PNIPAM peaks is due to the macrophase separation between the homopolymers of the blend and the low ratio of PNIPAM in the blend (25%), which prohibits the detection of the corresponding Raman peaks. In Fig. 6a, the Raman spectra of the isolated dewettted areas of the 50/50 PS/PNIPAM film show only the overtone (ca. 963 cm$^{-1}$) of the silicon optical phonon peak (520.5 cm$^{-1}$), confirming the nature of dewetting. The PS/PNIPAM film is spatially dewetted on the silicon substrate, resulting in the formation of isolated dewetted areas, due to the different wetting tendencies of PS and PNIPAM on silicon.[35] Figure 6b shows Raman spectra on three different locations of the PS/PNIPAM network that correspond to the vibrational modes of C-H for PS and PNIPAM. The locations of the Raman spectra are indicated by the corresponding stars in the optical micrographs. Additional to the C-H vibrational modes for PS, the C-H vibrational modes for PNIPAM appear at 2874 cm$^{-1}$, 2921 cm$^{-1}$, 2935 cm$^{-1}$, and 2976 cm$^{-1}$,[36] also verified by Raman spectroscopy on solid PNIPAM (Figs. S1 and S2 in Supplementary Information). Comparing the Raman spectra shown in Fig. 6b, spectrum # 1 presents peaks that correspond only to PS, indicating the absence of PNIPAM in this area, in contrast with spectrum #2 that presents the peaks of PNIPAM and a very weak PS peak at 3054 cm$^{-1}$, indicating mainly the presence of PNIPAM in this area. Additionally, spectrum #3 presents peaks that correspond to both PS and PNIPAM. Comparing all three spectra, we conclude that the 50/50 PS/PNIPAM film presents macrophase separation between the homopolymers of the blend.

Figure 7 shows water contact angle measurements of PS/PNIPAM films of 75/25 and 50/50 blend ratios, as well as pure PS, at 25 °C and 45 °C. We observe that the pure PS film is hydrophobic, with a contact angle ~100°, in contrast to the blend PS/PNIPAM films, which are hydrophilic. Due to the hydrophobicity of PS, the PS/PNIPAM films are less hydrophilic as the ratio of PS increases at 25°C.[12]

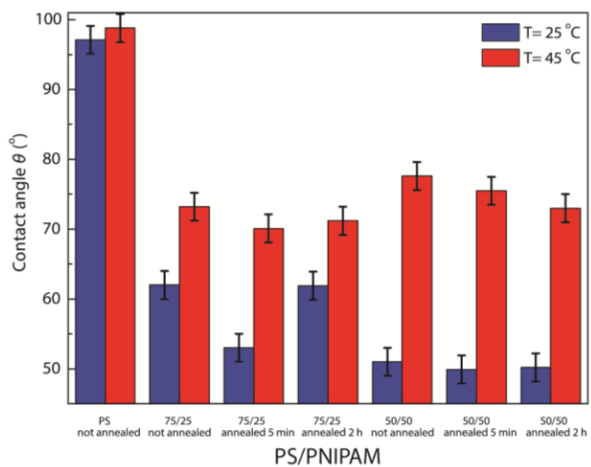

FIGURE 7. Bar graph of water contact angles, measured at 25 °C and 45 °C, on a PS film and on PS/PNIPAM films of 75/25 and 50/50 blend ratios (not annealed, annealed at 100 °C for 5 min, and annealed at 100 °C for 2 hours).

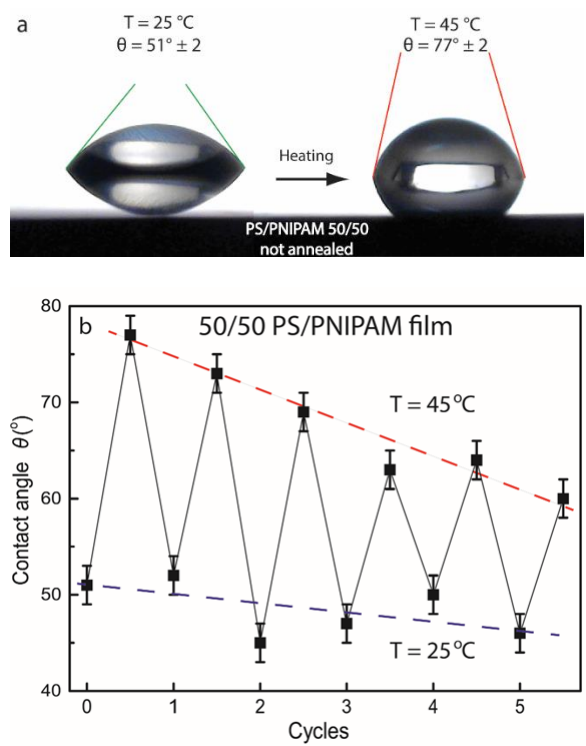

FIGURE 8. Water contact angle measurements on a PS/PNIPAM 50/50 film (a) at 25 °C and 45 °C and (b) for five cycles of heating/cooling at 45 °C and 25 °C. Dashed lines are guides to the eye for the trend of the contact angles at 25 °C (blue color) and 45 °C (red color).

Comparing Figs. 3 and 7 we observe that the homopolymer blend films PS/PNIPAM are more hydrophilic at environmental temperature (25 °C) than the diblock copolymer blend films PS/PS-b-PNIPAM. In contrast to the films of PS/PS-b-PNIPAM blends, the effect of switchable wettability of PNIPAM below and above the LCST is clearly observed in the films of the homopolymer blends PS/PNIPAM. The effect is more pronounced for the films of high PNIPAM ratio (50/50). The films of 75/25 and 50/50 ratios show an of increase their water contact angle by ~10° and ~25° between 25 °C and 45 °C, respectively. A contact angle measurement is shown in Figure 8a, which demonstrates an increase of 26° in the angle of contact between water and a 50/50 PS/PNIPAM film, after heating the film above the LCST of PNIPAM. Additionally, Figure 8b shows the reversible switching behaviour of a 50/50 PS/PNIPAM film, achieving reversible thermoresponsivity for five cycles of heating/cooling. We observe that the response of the 50/50 PS/PNIPAM film is higher in the beginning of the heating/cooling cycles than in the end, however, it is clearly observed that after five cycles the film still responds reversibly to the stimulus of temperature.

Annealing at 100 °C, either for 5 min or for 2 hours, does not affect significantly the switchable wetting behaviour of the films, as shown in Fig. 7. To understand the effect of annealing on the morphology of the homopolymer blend films, we monitor *in situ* the annealing process on a 75/25 and a 50/50 PS/PNIPAM film at 100 °C for 2 hours, using an optical microscope with a thermal control stage. As it was observed in Fig. 4, the PS/PNIPAM films show micrometric morphological features, therefore a change in their morphology upon annealing can be resolved by optical microscopy. Figures 9a (i)-(ix) and b (i)-(ix) show the evolution of the surface of a 75/25 and a 50/50 PS/PNIPAM film, respectively, during the annealing process. The images at t = 0 depict the films directly after the spin coating process, still at room temperature, and the temperature is raised to 100 °C at t = 2 min. Figures 9a (ix) and b (ix) correspond to the end of the 2-hour annealing process, when the films return to room temperature. We observe that there is no difference at the micro-morphology of the film surfaces before, during, and after the annealing process, which explains the similarity in the wetting behaviour between the annealed and the not annealed films.

Comparing the wetting behaviour of the homopolymer PS/PNIPAM films to that of the block copolymer PS/PS-b-PNIPAM films, we observe that the former demonstrate thermoresponsive wettability, presenting a systematic increase of their contact angles, as it is expected due to the PNIPAM characteristic property, while the latter do not respond systematically to the stimulus of temperature even if there are small changes upon heating. However, other studies of PS-b-PNIPAM copolymer brushes grafted onto flat silicon substrates by ATRP show an increase of the water contact angle as a function of temperature.[12,17] The grafting of PNIPAM on the PS layer results in the formation of copolymer PS-b-PNIPAM brushes with PNIPAM chains arranged upwards, free to form bonds with water molecules. On the other hand, with the spin coating technique employed here, the polymers adopt the thermodynamically favored structure, [26] which, in the case of PS-b-PNIPAM, corresponds to the chains of PNIPAM enriching the area near the bottom of the film, as was explained in Section 3.1. A similar behaviour has been observed with spin-casted PS-b-PMMA films on silicon, where the PMMA block was deposited preferentially on the silicon substrate due to its lower wetting energy.[29] On the other hand, our results on the wettability of the homopolymer PS/PNIPAM blend films agree with studies, which graft the PNIPAM monomer on silicon surfaces and measure a ~30° change of the water contact angle below and above the LCST of PNIPAM.[7,11,37,38]

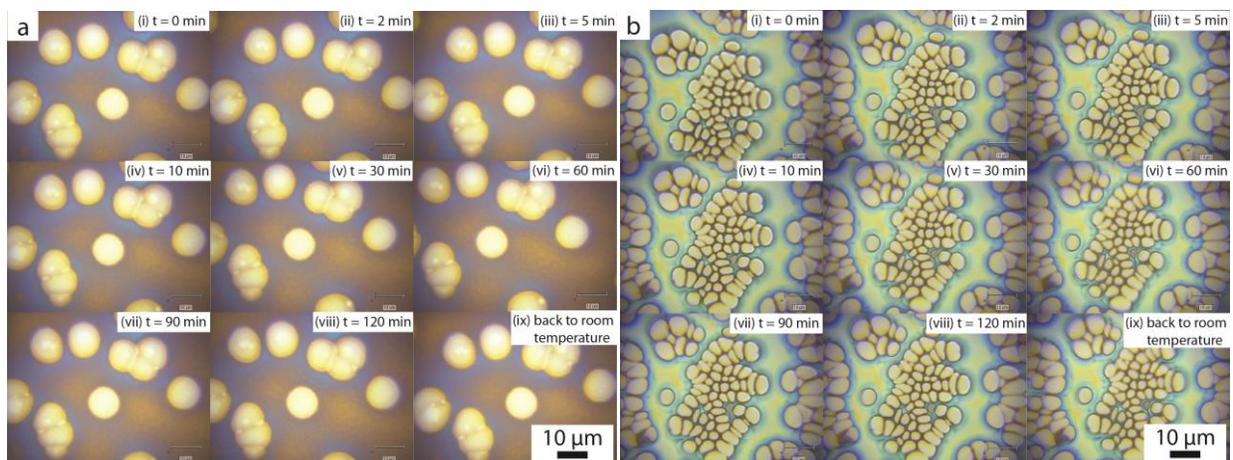

FIGURE 9. Optical microscope images of a (a) 75/25 and (b) 50/50 PS/PNIPAM film at different times during the 2-hour annealing process at 100 °C.

Spin-casted films of homopolymer blends are able to alter their wettability reversibly with temperature, even if they are spatially dewetted, because the PNIPAM chains are free on the film surface to interact with water molecules. Therefore, the spin coating technique is more suitable for the homopolymer blend films of PS/PNIPAM than the PS/PS-b-PNIPAM films, allowing thin films of homopolymer blends to demonstrate reversibly a switchable, thermoresponsive wetting behavior.

**CONCLUSIONS**

We develop thin polymer films by blending PS with PNIPAM and its diblock copolymer PS-b-PNIPAM, via spin coating on silicon substrates, and we study the morphology and wetting behaviour of the blend films with different compositions and drying conditions, achieving tunable wettability. The morphology of the films varies significantly with the blend ratio and the drying conditions. Blending polymers allows the formation of complex morphologies inducing micro- and nano-patterning on the film surface. Thermal annealing modifies the morphology of the film surfaces and the annealed films present smoother surfaces compared with the not annealed films. Concerning the film stability, films of both kinds of blends are stable after the water contact angle measurements. For the PS/PS-b-PNIPAM films, we do not observe an increase of the water contact angle with temperature, irrespectively of the blend ratio and drying conditions, because the thermoresponsive PNIPAM chains enrich preferentially the lower area of the film, close to the silicon substrate, and the non-thermoresponsive PS chains enrich the upper layer of the film, close to the surface. On the other hand, the films of homopolymer PS/PNIPAM blends show an increase of the water contact angle as high as 25° above the LCST of PNIPAM, demonstrating reversibly a tunable, thermoresponsive wetting behavior. Films of homopolymer blends with a higher ratio of PNIPAM exhibit bigger changes in their contact angles above the LCST. Concerning the drying conditions, thermal annealing does not affect the wetting behaviour of both kinds of blend films. In contrast to grafting techniques, which require elaborate equipment and metal catalysts, spin coating is a simple and cost-effective process for the development of thin films. To the best of our knowledge, there are no studies on the thermoresponsive wetting behaviour of spin-casted films of PS/PS-b-PNIPAM and PS/PNIPAM blends. Our results on the wettability of thin films of homopolymer blends provide the possibility to develop smart surfaces by simple and controllable methods, suitable for numerous applications that require control of wettability over large areas, such as self-cleaning surfaces, industrial surface coatings, tissue engineering, cell encapsulation, enzyme immobilization, sensing, and microfluidics, among others.

**ACKNOWLEDGEMENTS**

We acknowledge support of this work by the project "Advanced Materials and Devices" (MIS 5002409) which is implemented under the "Action for the Strategic Development on the Research and Technological Sector", funded by the Operational Programme "Competitiveness, Entrepreneurship and Innovation" (NSRF 2014-2020) and co-financed by Greece and the European Union (European Regional Development Fund). M. Kanidi acknowledges support through a Ph.D. fellowship by the General Secretariat for Research and Technology (GSRT) and the Hellenic Foundation for Research and Innovation (HFRI). We also thank Dr. G. D. Chryssikos and Dr. E. Siranidi for their help with Raman spectroscopy measurements and Dr. C. Chochos for his help with SEM measurements.

**REFERENCES**

1. T. Sun, L. Feng, X. Gao, L. Jiang, *Accounts Chem. Res.*, **2005**, 38, 644.
2. M. Ma, R. M. Hill, *Curr. Opin. Colloid In.*, **2006**, 11, 193.
3. F. Xia, L. Jiang, *Adv. Mat.*, **2008**, 20, 2842.
4. F. Guo, Z. Guo, *RSC Adv.*, **2016**, 6, 36623.
5. B. Xin, J. Hao, *Chem. Soc. Rev.*, **2010**, 39, 769.
6. G. H. Schild, *Prog. Polym. Sci.*, **1992**, 17, 163.
7. T. Sun, G. Wang, L. Feng, B. Liu, Y. Ma, L. Jiang, D. Zhu, *Angew. Chem. Int. Edit.*, **2004**, 43, 357.
8. C. T. Schwall, I. A. Banerjee, Materials, **2009**, 2, 577.
9. O. Sedlacek, P. Cernoch, J. Kucka, R. Konefal, P. Stepanek, M. Vetrik, T. P. Lodge, M. Hruby, *Langmuir*, **2016**, 32, 6115.
10. T. Meng, R. Xie, Y. C. Chen, C. J. Cheng, P. F. Li, X. J. Ju, L. Y. Chu, *J. Membrane Sci.*, **2010**, 349, 258.
11. L. Li, Y. Zhu, B. Li, C. Gao, *Langmuir*, **2008**, 24, 13632.
12. Q. Yu, Y. Zhang, H. Chen, F. Zhou, Z. Wu, H. Huang, J. L. Brash, *Langmuir*, **2010**, 26, 8582.
13. I. Luzinov, S. Minko, V. V. Tsukruk, *Soft Matter*, **2008**, 4, 714.
14. T. Yakushiji, K. Sakai, *Langmuir*, **1998**, 14, 4657.
15. Y. G. Takei, T. Aoki, K. Sanui, N. Ogata, Y. Sakurai, T. Okano, *Macromolecules*, **1994**, 27, 6163.
16. S. Kumar, Y. L. Dory, M. Lepage, Y. Zhao, *Macromolecules*, **2011**, 44, 7385.
17. J. K. Chen, J. H. Wang, S. K. Fan, J. Y. Chang, *J. Phys. Chem. C*, **2012**, 116, 6980.
18. P. W. Majewski, K. G. Yager, *J. Phys.: Condens. Mat.*, **2016**, 28, 403002.
19. E. Stamm, Polymer Surfaces and Interfaces, Characterization, Modification and Applications, Springer, 2008.
20. F. S. Bates, G. H. Fredrickson, *Annu. Rev. Phys. Chem.*, **1990**, 41, 525.
21. F. S. Bates, *Science*, **1991**, 251, 898-905.
22. Y. Huang, H. Liu, Y. Hu, *Macromol. Theor. Simul.*, **2006**, 15, 321.
23. M. Rosales-Guzman, R. Alexander-Katz, P. Castillo-Ocampo, A. Vega-Rios, A. Licea-Claverie, *J. Polym. Sci., Part B: Polym. Phys.*, **2013**, 51, 1368.
24. L. Cui, Y. Ding, X. Li, Z. Wang, Y. Han, *Thin Solid Films*, **2006**, 515, 2038.
25. D. A. Winesett, H. Ade, J. Sokolov, M. Rafailovich, S. Zhu, *Polym. Int.*, **2000**, 49, 458.
26. K. Norrman, A. Ghanbari-Siahkali, N. B. Larsen, *Annu. Rep. Prog. Chem. C*, **2005**, 101, 174.

27. W. Wang, E. Metwalli, J. Perlich, C. M. Papadakis, R. Cubitt, P. Muller-Buschbaum, *Macromolecules*, **2009**, 42, 9041.
28. Q. Yang, K. Loos, *Polymers*, **2017**, 9, 525.
29. X. Zhang, F. J. Douglas, L. R. Jones, *Soft Matter*, 2012, 8, 4980.
30. E. Huang, S. Pruzinsky, T. P. Russel, J. Mays, C. J. Hawker, *Macromolecules*, **1999**, 32, 5299.
31. N. Bhandaru, A. Das, N. Salunke, R. Mukherjee, *Nano Lett.*, **2014**, 14, 7009.
32. N. Bhandaru, A. Karim, R. Mukherjee, Soft Matter, **2017**, 13, 4709.
33. M. Geoghegan, G. Krausch, *Prog. Polym. Sci.*, **2003**, 28, 261.
34. W. M. Sears, J. L. Hunt, J. R. Stevens, J. Chem. Phys., **1981**, 75, 1589.
35. D. U. Ahn, Z. Wang, I. P. Campbell, M. P. Stoykovich, Y. Ding, *Polymer*, **2012**, 53, 4187.
36. J. Dybal, M. Trchova, P. Schmidt, *Vib. Spectrosc.*, **2009**, 51, 44.
37. Q. He, A. Kuller, M. Grunze, J. Li, *Langmuir*, **2007**, 23, 3981.
38. J. K. Chen, C. J. Chang, *Materials*, **2014**, 7, 805.

**GRAPHICAL ABSTRACT**

Maria Kanidi, Aristeidis Papagiannopoulos, Athanasios Skandalis, Maria Kandyla*, and Stergios Pispas

Thin films of PS/PS-b-PNIPAM and PS/PNIPAM polymer blends with tunable wettability

Spin-casted polymeric films prepared by blends of PS/PNIPAM and PS/PS-b-PNIPAM, using different composition and drying conditions, show tunable wetting behaviour upon heating. Films of PS/PNIPAM and PS/PS-b-PNIPAM blends exhibit different surface morphology due to the coexistence of phase separation and dewetting phenomena. Studying the wetting behaviour of the films, the PS/PNIPAM films demonstrate reversible thermoresponsive wettability, while the PS/PS-b-PNIPAM films do not respond systematically to the stimulus of temperature.

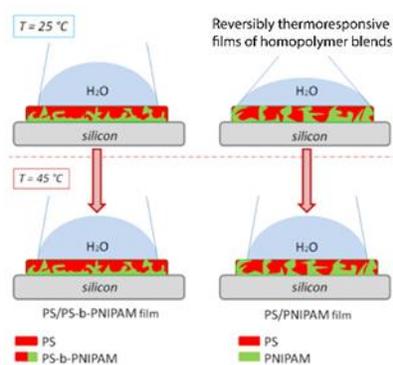